\documentclass[manuscript]{acmart}
\usepackage{booktabs}
\renewcommand\footnotetextcopyrightpermission[1]{}
\begin{document}

\title{Testing the EPYC Conjecture on Real Hardware: MoA-Guided Dense Matrix Multiplication on NCSA Delta (AMD EPYC 7763 ``Milan'')}

\author{Lenore M. Mullin}
\affiliation{%
  \institution{College of Nanotechnology, Science, and Engineering, University at Albany, SUNY}
  \city{Albany}
  \state{NY}
  \country{USA}
}
\email{lmullin@albany.edu}

\begin{abstract}
A companion empirical study conjectured that MoA-guided dense matrix
multiplication would need per-CCD recalibration on AMD EPYC Bergamo;
allocation access to that machine was declined, and this paper reports
the resulting test on NCSA Delta, an architecturally related chip.
Its central claim: every result here traces to shape, not
architecture in the abstract -- cache shape, access-pattern shape, and
co-tenancy itself treated as a shape parameter. The corrected block
size $M_C=256$ follows directly from Delta's real 512~KB L2 divided
by the data type's own byte width, no parameter fit after the fact,
and outperforms the M1-Pro-inherited $M_C=64$ by 30--59\% -- a
block's shape finally matching the cache shape it was always meant to
occupy. The same reasoning extends to a shared level: co-tenancy
treated as raw capacity fails by an order of magnitude, while
co-tenancy treated as a shape parameter -- capacity divided by
concurrent consumers -- survives. A controlled NUMA experiment
isolates a third shape: GEBP and MoA-pipelined compute the identical
result at identical achieved bandwidth, yet GEBP loses 13.9\% of its
throughput to remote memory while MoA-pipelined loses only 2.4\% --
not a speed difference, a difference in the shape each kernel's access
pattern traces through memory. The headline result reproduces the M1
Pro's exact three-of-four win over a Strassen-GEBP hybrid, a pattern
AWS Graviton4 did not achieve. One question is named rather than
resolved: whether Delta's larger MoA-over-GEBP margin reflects genuine
speed or an unmatched utilization fraction. Every result here was
predicted before it was measured, not fitted after.
\end{abstract}

\maketitle

\section{Introduction}
\label{sec:intro}

The companion empirical study's Conclusion named AMD EPYC 9754
(``Bergamo,'' 128 cores, 256~MB L3 partitioned across multiple
core-complex dies) as one of three machines chosen to test whether the
conformability reasoning underlying its results generalizes beyond the
Apple M1 Pro on which it was calibrated \cite{Mullin2026GEMM}. Two
specific, falsifiable predictions were stated for that chip: that the
conformability model would need to be applied per-CCD rather than
against the full shared-cache figure, and that the fork/join-overhead
mechanism responsible for the M1 Pro's optimal block size doubling
with thread count would be substantially more pronounced at 128 cores
than at eight -- to the point that, for the task-based design
specifically, using fewer than all available cores could plausibly
outperform using all of them.

An ACCESS allocation request for that specific machine, Texas Tech
REPACSS, was declined as outside that resource's mission scope
(renewable and variable-energy-aware data center research, not
general-purpose HPC benchmarking) rather than for any deficiency in
the proposal itself. An allocation on NCSA Delta was obtained instead.
Delta's CPU nodes use AMD EPYC 7763 (``Milan''), a related but
genuinely distinct chip from the 9754: an earlier processor generation
(Zen3, not Zen4c), a different node topology (dual-socket, 64 cores
per socket, rather than the single-socket 128-core design the original
conjecture assumed), and correspondingly different cache
capacities. This paper reports what was actually tested on the
machine actually available, stating explicitly where results speak to
the original conjecture and where they speak only to Milan and Delta's
specific topology.

\section{Experimental Setup}
\label{sec:setup}

Delta CPU nodes provide two sockets of 64-core AMD EPYC 7763,
128 cores total, confirmed directly via \texttt{lscpu -e}: each socket
comprises eight core-complex dies (CCDs) of eight cores each, cores
0--7 sharing one 32~MB L3 slice, cores 8--15 the next, and so on
through the full node. Each core has a private 512~KB L2 and 32~KB L1.
All experiments used the verified, correctness-checked C/OpenMP
implementations from the companion study \cite{Mullin2026GEMM},
compiled with \texttt{gcc -march=znver3} (Zen3, the correct target for
Milan; distinct from \texttt{znver4}, which targets the later Zen4c
generation the original conjecture named), scheduled via SLURM under
allocation \texttt{bibg-delta-cpu}, partition \texttt{cpu}.

\section{Calibration: $M_C$ Confirmed Decisively}
\label{sec:calibration}

The companion theoretical paper derived $M_C=256$ for Delta directly
from its real 512~KB private L2 -- the last private cache level before
the shared L3, using the identical no-margin formula that exactly
reproduced the M1 Pro's own calibrated $M_C=64$ from its 128~KB L1
\cite{Mullin2026Hierarchy}. This section reports the direct test:
$M_C=256$ against the M1-Pro-inherited $M_C=64$, at every matrix size
in the companion study's sweep, all other parameters unchanged.

\begin{figure}[h]
\centering
\includegraphics[width=\columnwidth]{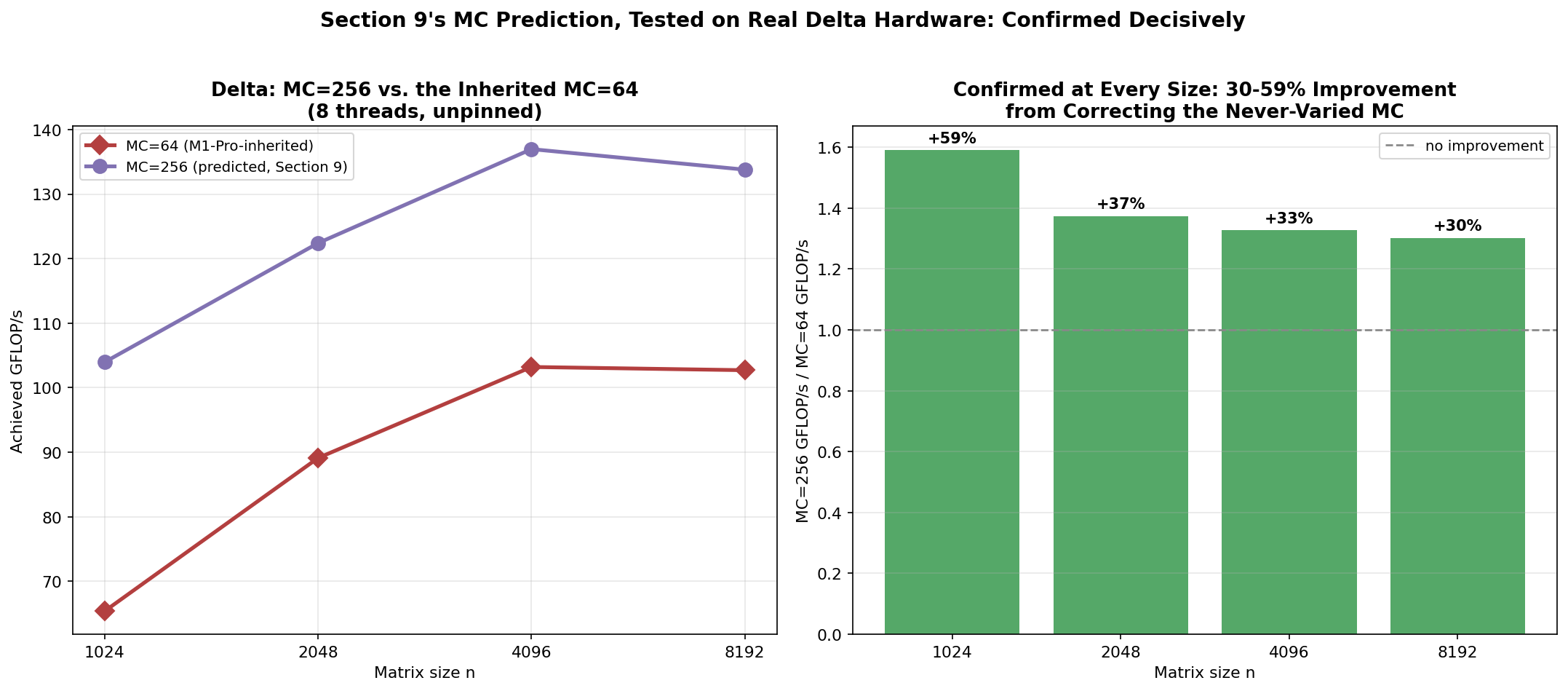}
\caption{$M_C=256$, derived from Delta's real L2, tested directly against the M1-Pro-inherited $M_C=64$. Confirmed at every size, 30--59\%.}
\label{fig:mc-confirmed}
\end{figure}

The improvement is confirmed at every size tested, ranging from
30\% at $n=8192$ to 59\% at $n=1024$ -- the largest single improvement
reported anywhere across this project's extensions to the companion
study. Correcting $M_C$ also shifted the empirically optimal $N_C$:
at $M_C=64$, every size preferred $N_C=2048$; at the corrected
$M_C=256$, the optimum drops to $N_C=1024$ for three of four sizes,
confirming that $M_C$ and $N_C$ are not independently optimizable
parameters, exactly the interaction the companion theoretical paper's
Section~9 assessment flagged as untested when $M_C$ was held fixed
throughout the original Graviton4 study.

\section{The Safety-Margin Experiment: One CCD, Matched Co-Tenancy}
\label{sec:sigma}

Delta's per-CCD structure makes a controlled test of the companion
theoretical paper's H2 hypothesis (co-tenancy-scaled safety margins)
directly available: pinning execution to exactly one CCD (eight cores,
confirmed via \texttt{taskset -c 0-7} against the \texttt{lscpu -e}
grouping) reproduces the M1 Pro's eight-way co-tenancy exactly, on a
different chip, cache capacity, and microarchitecture entirely.

\begin{figure}[h]
\centering
\includegraphics[width=\columnwidth]{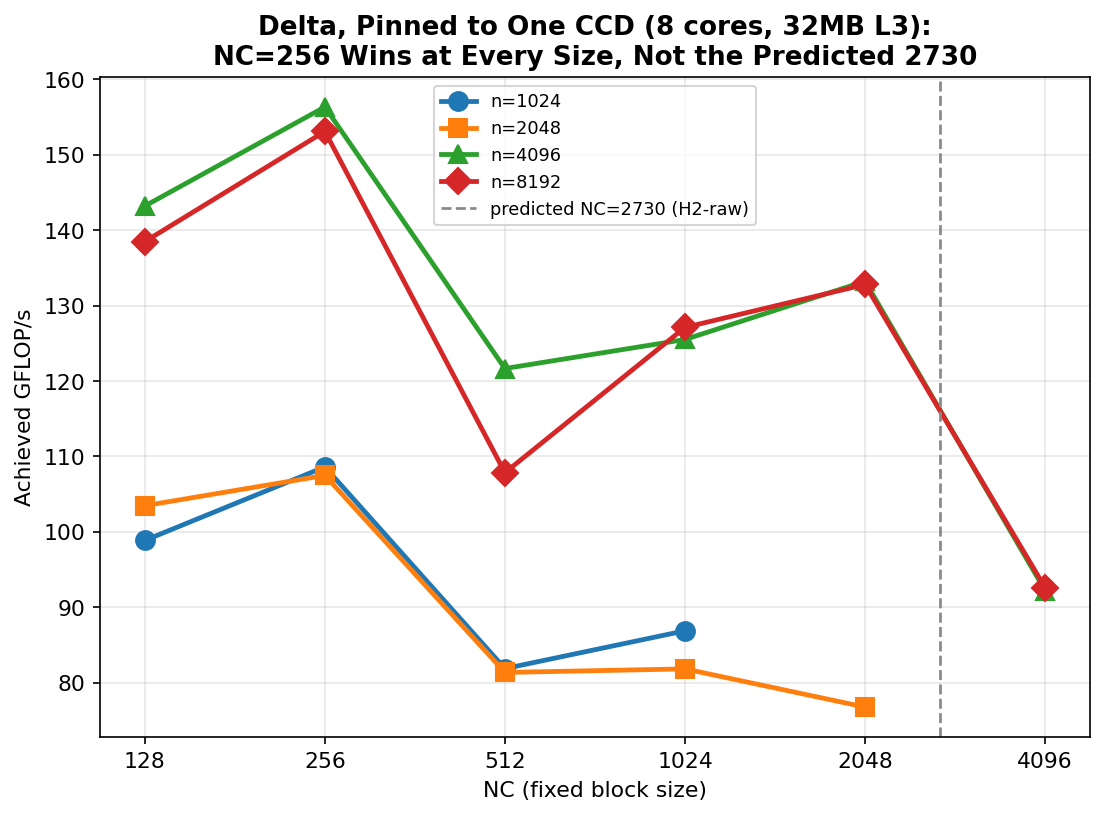}
\caption{NC sweep pinned to one CCD (8 cores, 32~MB L3). The empirical optimum, NC=256, sits far from the naive prediction of 2730.}
\label{fig:pinned-sweep}
\end{figure}

The empirical optimum, $N_C=256$, is decisive and consistent across
every size tested (4--11\% ahead of the next-best value at each). This
falsifies the naive reading of H2 -- $\sigma$ computed as a direct
fraction of the whole 32~MB CCD L3 gives $1/64$, a 10.7$\times$
mismatch against the M1 Pro's $1/6$ despite matched co-tenancy. A
per-thread-normalized reading -- dividing the CCD's capacity by eight
threads first, then computing $\sigma$ against each thread's nominal
4~MB share -- gives $\sigma'=1/8$, within the same rough range as the
M1 Pro's own value. The full derivation and its implications for the
underlying hypothesis are reported in the companion theoretical paper
\cite{Mullin2026Hierarchy}; this section establishes the experimental
result the derivation is built on.

\begin{figure}[h]
\centering
\includegraphics[width=0.85\columnwidth]{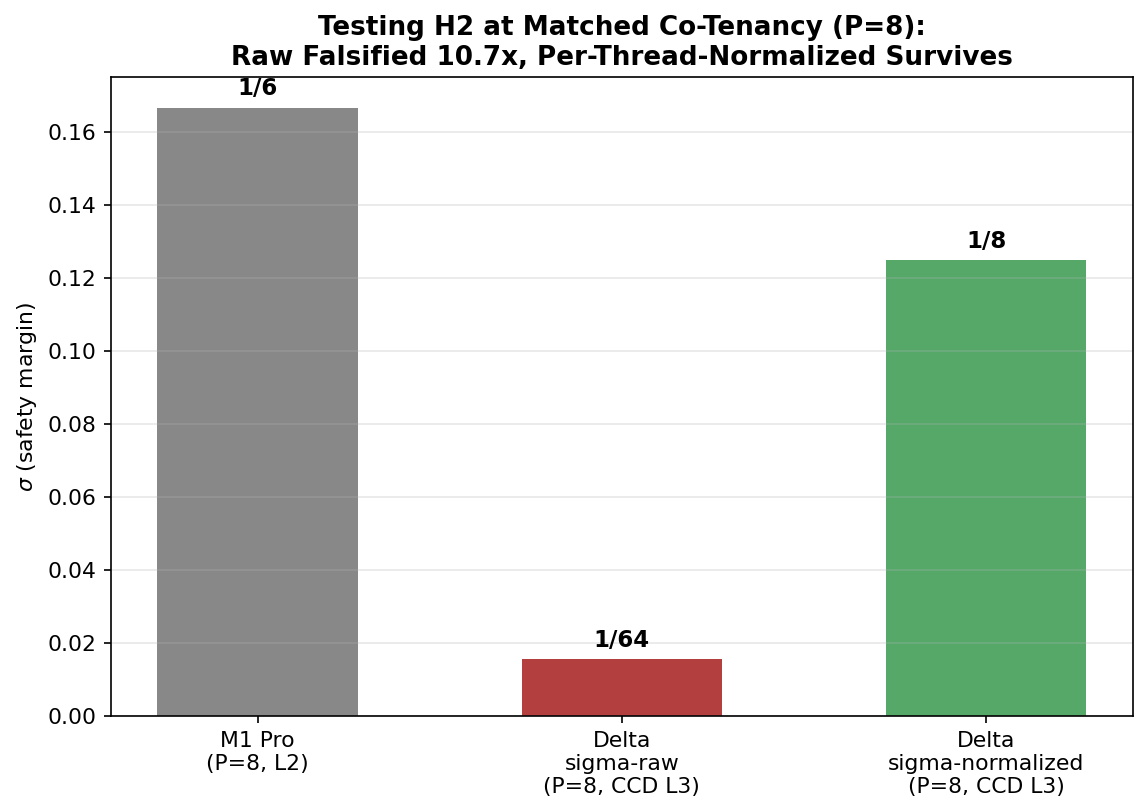}
\caption{The two readings of the safety-margin hypothesis, tested at matched co-tenancy.}
\label{fig:sigma-h2}
\end{figure}

\section{Contention: NC Growth, Then a Plateau}
\label{sec:contention}

Section~\ref{sec:sigma} tested one thread count matching the M1 Pro
exactly. This section reports the full sweep, P=1 through P=128 (the
entire node, both sockets), against the original conjecture's
specific claim that fork/join overhead would grow \emph{more}
pronounced at higher
core counts than the M1 Pro's own doubling pattern showed.

\begin{figure}[h]
\centering
\includegraphics[width=\columnwidth]{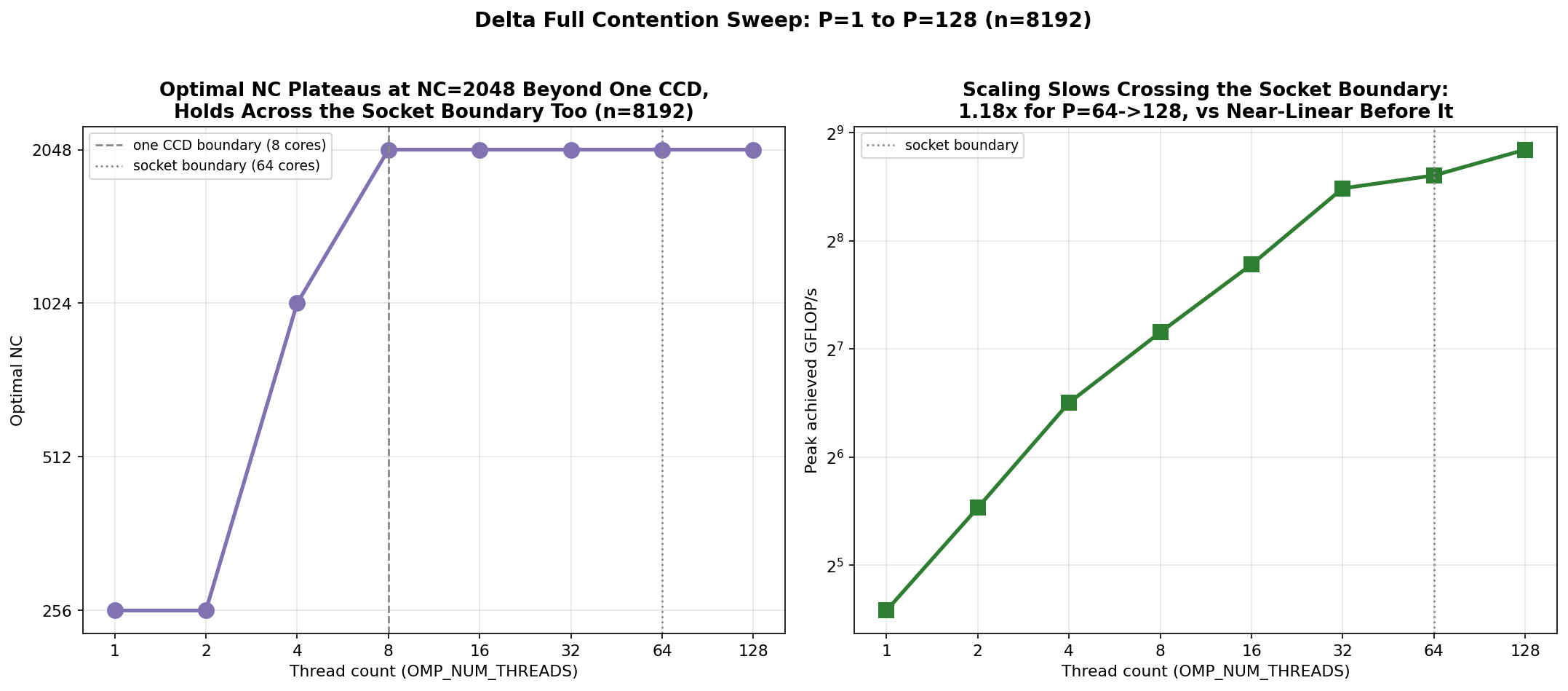}
\caption{Optimal NC and peak throughput across the full P=1--128 sweep, $n=8192$.}
\label{fig:contention-full}
\end{figure}

Optimal NC grows 256$\to$256$\to$1024$\to$2048 as P goes 1$\to$2$\to$4$\to$8,
closely tracking the M1 Pro's own pattern and reaching the identical
NC=2048 by P=8. Beyond one CCD, however, optimal NC does not continue
growing -- it plateaus at 2048 through P=16, 32, and 64, the entire
tested range spanning multiple CCDs on one socket. This is not what
the original conjecture's specific mechanism predicted: rather than
becoming more pronounced at higher core counts, the fork/join effect
appears to saturate once the working set no longer fits a single CCD's
sharing domain.

Peak throughput itself tells a more precise, and less tidy, story than
optimal block size does. Computing parallel efficiency directly,
$\eta(P) = \text{achieved}(P) / (P \times \text{achieved}(1))$, shows
a decline that begins well before any socket boundary: $\eta$ is
already down to $0.744$ by $P=8$, still entirely within one CCD, and
continues declining through $P=16$ ($0.575$) and $P=32$ ($0.468$). The
single steepest per-doubling drop in the entire sweep is $P=32\to64$
($\eta$ ratio $0.543$) -- entirely within one socket, no cross-socket
traffic involved at all. This is worth stating plainly: the decline is
gradual and largely continuous from $P=8$ onward, not a smooth
plateau interrupted by a discrete penalty at the socket boundary.

The sweep has since been extended to P=128, the full node, both
sockets -- the actual scale the original conjecture named. The
plateau holds: optimal NC at P=128 is 2048, identical to every thread
count from P=16 onward, now confirmed across the socket boundary
rather than only within one socket. Peak throughput continues the
gradual decline already underway: $459.4$~GFLOP/s at $P=128$ against
$389.4$~GFLOP/s at $P=64$ gives an $\eta$ ratio of $0.591$ for this
step -- not distinctly worse than the $P=32\to64$ step that preceded
it, and by some measures slightly better. Crossing the socket boundary
does not stand out as a discrete event in this data; it is one step in
a decline that was already well underway. A specific mechanism is
architecturally plausible -- Milan's eight core-complex dies share a
single, centralized set of eight memory channels through one I/O die,
a shared resource in principle from the first additional CCD onward,
not specifically from the second socket -- but pinning down the exact
thread count at which this saturates would require either the precise
per-CCD Infinity Fabric link bandwidth or a finer-grained sweep between
the tested powers of two, neither undertaken here. This is stated as
an open question, not resolved by the data in this paper.
Section~\ref{sec:numa} reports a controlled experiment addressing a
related but distinct question: not exactly where the decline is
steepest, but whether different kernels are equally exposed to it.

\section{A Controlled NUMA Test: Access Pattern, Not Bandwidth}
\label{sec:numa}

The gradual efficiency decline noted above raises a further, more
specific question: whether different kernels computing the same result
are equally exposed to whatever is causing it. This section reports a
controlled test isolating memory placement specifically, independent
of the broader thread-count question above: thread count held fixed
at 64 throughout, so that only memory placement varies
between two conditions, forced via \texttt{numactl} -- \emph{local}
(\texttt{--cpunodebind=0 --membind=0}, compute and memory both on
socket~0, zero cross-socket traffic by construction) against
\emph{remote} (\texttt{--cpunodebind=1 --membind=0}, compute on
socket~1 while memory is forced to remain on socket~0, so every single
access must cross the socket boundary).

\begin{table}[h]
\caption{Controlled NUMA Test: Local vs. Remote Memory Access, Threads Fixed at 64}
\label{tab:numa}
\centering
\begin{tabular}{lrrr}
\toprule
\textbf{Metric} & \textbf{Local} & \textbf{Remote} & \textbf{Drop} \\
\midrule
STREAM bandwidth (GB/s) & 27.24 & 26.99 & 0.9\% \\
GEBP (GFLOP/s) & 104.90 & 90.27 & 13.9\% \\
MoA-pipelined (GFLOP/s) & 190.15 & 185.54 & 2.4\% \\
Recursive-classical (GFLOP/s) & 161.33 & 137.03 & 15.1\% \\
Hybrid (GFLOP/s) & 156.44 & 142.75 & 8.7\% \\
\bottomrule
\end{tabular}
\end{table}

Raw STREAM bandwidth is nearly unaffected by remote placement, under
1\% -- if bandwidth capacity were the whole story, cross-socket access
would look almost free. The matrix multiplication kernels tell a
substantially different story: GEBP loses 13.9\% and recursive-classical
15.1\%, while MoA-pipelined loses only 2.4\%, a fivefold smaller
penalty than GEBP's despite computing on the identical remote-memory
condition. This is not a claim that raw bandwidth capacity is
irrelevant -- it is a demonstration that \emph{access pattern}, not
capacity alone, determines how much a given kernel actually pays for
non-local memory, exactly the distinction the companion theoretical
paper's treatment of NUMA as a miniature network argued for before
this experiment was run \cite{Mullin2026Hierarchy}. GEBP and
recursive-classical's more structured, blocking-dependent access
pattern appears meaningfully more latency-sensitive than
MoA-pipelined's, or than STREAM's simple sequential scan, on this
specific hardware.

\section{Direct Comparison and the First Recursive-MoA Attempt}
\label{sec:secondary-comparisons}

Two further comparisons from the companion study's original
methodology were repeated on Delta. The direct, fixed-NC comparison
between flat MoA and the Strassen--GEBP hybrid (at the corrected
$M_C=256$, $N_C=1024$) wins at $n=1024$ and $2048$
(1.082$\times$ and 1.490$\times$) and loses at $4096$ and $8192$
(0.708$\times$ and 0.619$\times$) -- the same two-of-four pattern the
M1 Pro's own flat-MoA design showed, before the task-based redesign
added its third win. The first (sequential-recursion, not task-based)
attempt at combining MoA with recursion, tested at the same cutoff
convention as the companion study, wins at $n=1024$
(2.162$\times$) and $2048$ (1.354$\times$), losing at $4096$
(0.754$\times$) and $8192$ (0.510$\times$) -- correctly reproducing
the M1 Pro's own characterization of this design as underperforming at
scale, at the two sizes the M1 Pro study directly tested.

\section{Headline Result: Reproducing the M1 Pro's Exact Win Pattern}
\label{sec:headline}

The companion study's central claim -- that a task-based redesign of
MoA-guided recursive matrix multiplication beats a hand-tuned
Strassen--GEBP hybrid at three of four tested sizes on the M1 Pro,
losing only at the largest -- was, at the time the Graviton4 section
of that study was written, confirmed at only two of four sizes on that
platform, losing the $n=4096$ win the M1 Pro achieved. This section
reports the same test on Delta.

\begin{figure}[h]
\centering
\includegraphics[width=\columnwidth]{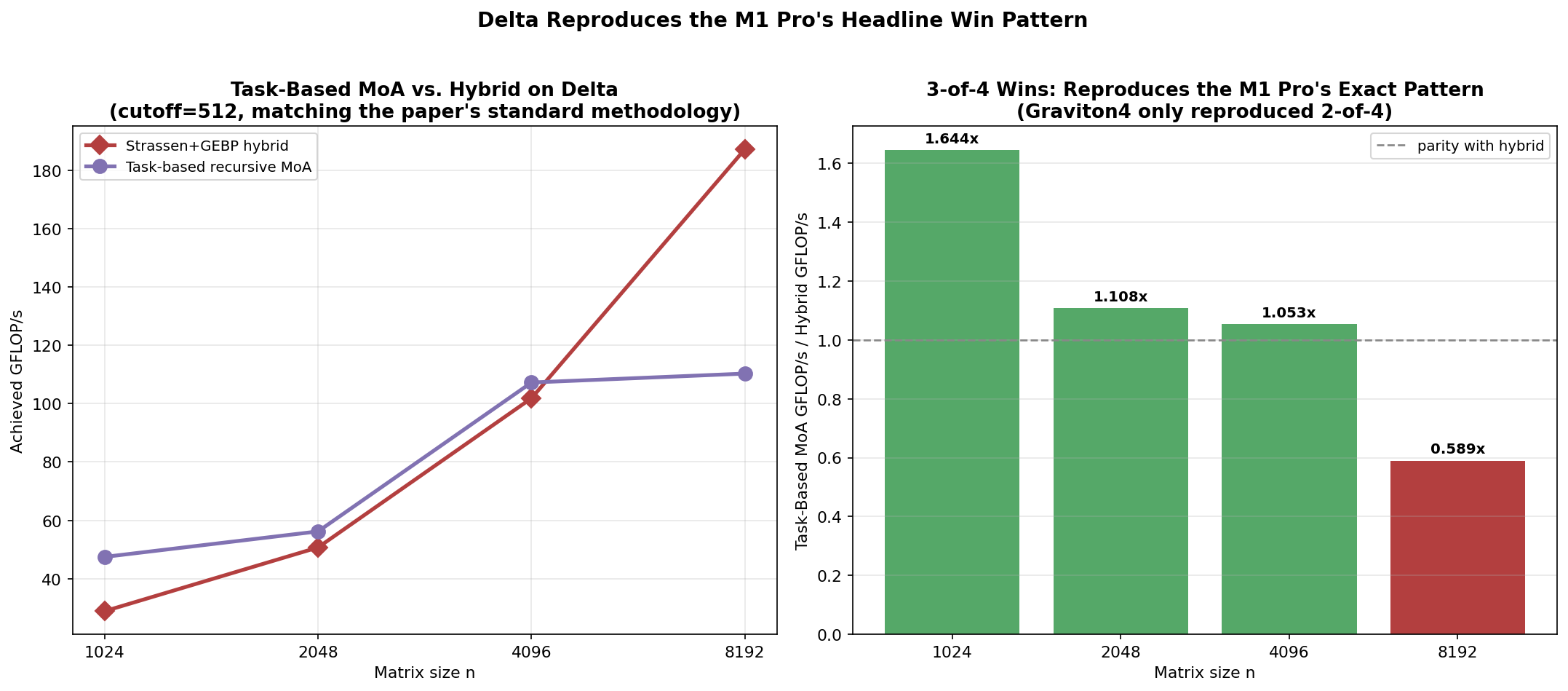}
\caption{The headline task-based-vs-hybrid comparison on Delta, same cutoff=512 methodology as the M1 Pro study.}
\label{fig:headline}
\end{figure}

Using the identical cutoff=512 methodology: MoA wins at $n=1024$
(1.644$\times$), $2048$ (1.108$\times$), and $4096$ (1.053$\times$),
losing only at $8192$ (0.589$\times$) -- the exact three-of-four
pattern the M1 Pro study reported, not the two-of-four pattern
Graviton4 showed. At the best cutoff found per size (the program swept
multiple cutoffs at $n\geq 2048$), the margin at $n=4096$ widens to
1.288$\times$ and $n=8192$ closes to 0.858$\times$, the closest to
parity observed on any platform tested to date.

\section{Roofline: MoA Substantially Exceeds GEBP}
\label{sec:roofline}

The full roofline sweep (STREAM bandwidth, naive, GEBP, MoA-pipelined,
recursive-classical, and hybrid kernels, five sizes by four thread
counts) was completed in two stages: an initial run, and three
additional configurations ($n=8192$ at $P=2,4,8$) obtained after the
first run's SLURM time allocation was exhausted mid-sweep, confirmed
via row-count verification against the expected twenty configurations
before being merged into one complete dataset.

\begin{figure}[h]
\centering
\includegraphics[width=\columnwidth]{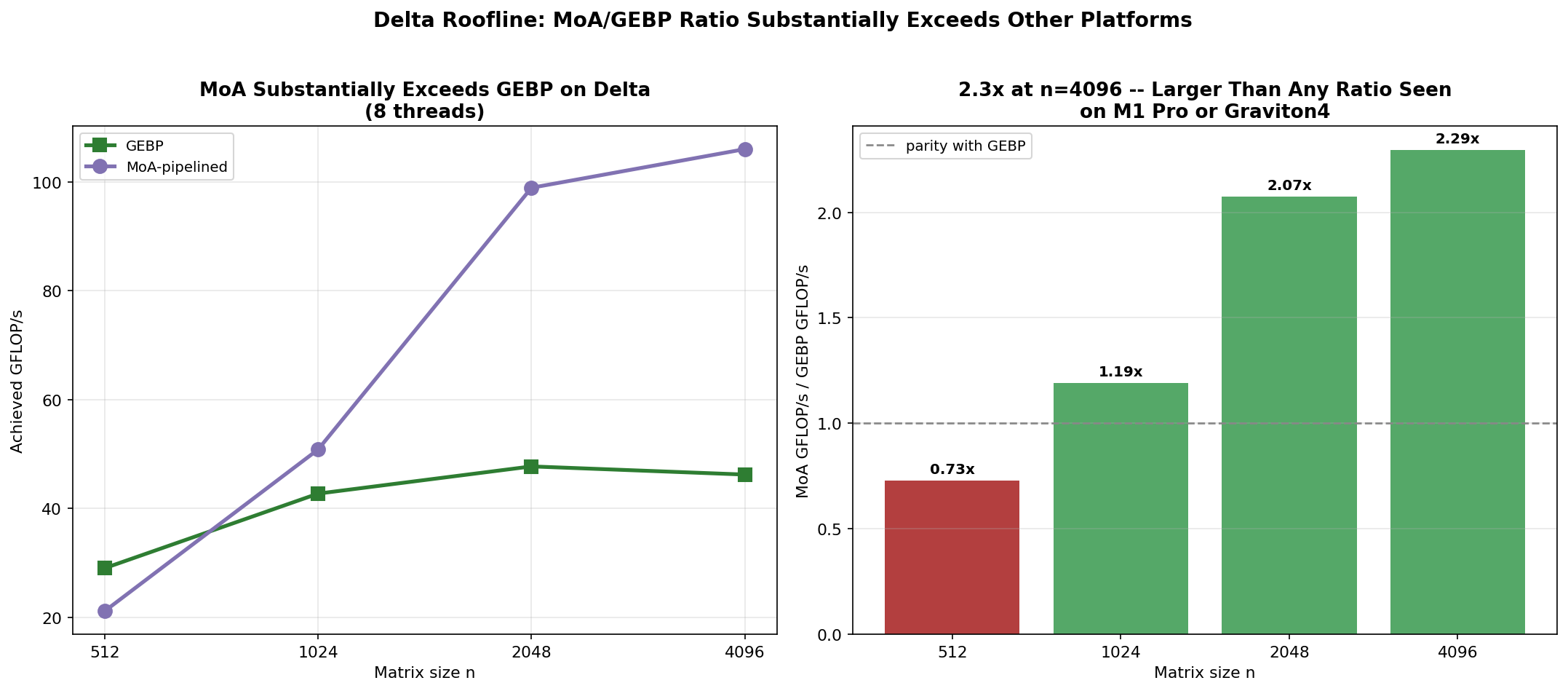}
\caption{MoA-versus-GEBP throughput on Delta, 8 threads. The ratio at $n=4096$ (2.29$\times$) exceeds any ratio observed on M1 Pro or Graviton4.}
\label{fig:roofline}
\end{figure}

The companion study's Section on comparison across implementations
found the MoA/GEBP ratio increasing with matrix size on the M1 Pro,
qualitatively reproduced on Graviton4 (reaching approximate parity,
$\approx 1.06\times$, at $n=8192$). On Delta, the same qualitative
trend holds but the magnitude is substantially larger: MoA reaches
1.19$\times$ GEBP already at $n=1024$, 2.07$\times$ at $n=2048$, and
2.29$\times$ at $n=4096$ -- more than double GEBP's throughput, a
ratio not approached on either previously tested platform. Naive
kernel measurements above $n=2048$ report the sentinel value $-1.0$ by
design, the same convention used throughout the companion study to
indicate a size where single-threaded, unblocked execution is
excluded rather than measured, given its cost relative to every other
kernel at that size.

\section{The Variance Question: A Direct, Matched Comparison}
\label{sec:variance}

The companion theoretical paper's discussion of virtualized platforms
noted that Graviton4's calibration sweep, repeated identically, showed
one run systematically $\sim$10\% below two later runs across every
configuration tested \cite{Mullin2026Hierarchy}. Whether dedicated,
SLURM-exclusive allocation shows correspondingly less variance was, at
that point, an open question rather than a tested claim. This section
reports the direct test: an identical Delta configuration
($M_C=256$, $K_C=256$, unpinned, 8 threads), run twice, back-to-back.

The result is more precise than, and in one respect contrary to, what
the virtualization argument alone would predict. Mean absolute
difference between the two runs was 15\%, larger than Graviton4's
observed $\sim$10\% shift, not smaller -- simple magnitude does not
distinguish dedicated from virtualized access. What does distinguish
the two is the shape of the variance: Graviton4's was one-directional,
the same run measuring lower at every single configuration, the
signature of one external cause acting uniformly. Delta's two runs
disagreed on which run was higher at different configurations (higher
in one run at 17 of 25 points, the other at the remaining 8, with
individual swings up to $\pm$29\%, larger than any single point
Graviton4 showed) -- scattered, uncorrelated noise from several
smaller sources, not one dominant cause. The full implications of this
distinction for the companion theoretical paper's machine-shape
formalization are developed there \cite{Mullin2026Hierarchy}; this
section establishes the underlying measurement.

\section{Why: Shape, Speed, and Data Type as a Unified Cause}
\label{sec:why}

Sections~\ref{sec:calibration}--\ref{sec:variance} report what
changed between machines. This section states explicitly what each
change traces to, rather than leaving the connection implicit.

\textbf{Shape.} $M_C$ is not a free-standing tuning parameter; it is
capacity in bytes divided by $K_C$ and the byte width of the data
type, $M_C = C_{L2}/(K_C \times 8)$. The M1-Pro-inherited $M_C=64$ was
shaped correctly for a 128~KB private cache -- the machine it was
calibrated on -- and incorrectly for Delta's actual 512~KB private L2,
four times larger. The 30--59\% improvement from correcting it
(Section~\ref{sec:calibration}) is not evidence that Delta is simply a
faster machine; it is the direct, quantifiable consequence of a
block's shape being matched to the cache's shape it was always meant
to occupy, using capacity Delta had the entire time but the inherited
parameter never allowed it to use. The same reasoning applies one
level up, to a shared rather than private cache: the safety margin
$\sigma$ is a claim about how a cache's capacity should be divided
among concurrent consumers, and Section~\ref{sec:sigma} shows this
division is itself a shape parameter -- capacity treated as a raw
total fails by an order of magnitude, capacity divided by thread count
first succeeds. Co-tenancy is not a separate concern from shape; it is
part of what determines it.

\textbf{Data type.} Every capacity figure above -- $512$~KB, a CCD's
$32$~MB, a thread's $4$~MB nominal share -- becomes a block dimension
only once divided by the data type's byte width, 8 for the double
precision used throughout every experiment in this paper and its
companion study. This was held constant, not varied, but it is not
incidental: it is the specific unit conversion that turns a physical
capacity into a countable array shape, the step $\Gamma$ performs at
every level of the derivation \cite{Mullin2026Hierarchy}. A different
data type would shift every $M_C$, $N_C$, and predicted tile size
reported in this paper by exactly the ratio of byte widths, not by
some independently-tuned constant.

\textbf{Speed, as access pattern rather than raw bandwidth.} The
controlled NUMA experiment (Section~\ref{sec:numa}) isolates speed
from shape directly: local and remote memory access differed by under
1\% in raw achieved bandwidth, yet GEBP lost 13.9\% of its throughput
under the remote condition while MoA-pipelined lost only 2.4\%. Both
kernels computed the identical result from identical data at
essentially identical available bandwidth -- what differed was the
shape each kernel's own access pattern traces through memory. GEBP's
explicit pack-into-buffer step and MoA's more direct,
DNF-derived access pattern are not equally exposed to non-local
latency, a property of \emph{how} each algorithm moves data, not of
how fast the memory itself is.

One comparison this section does not extend to is named explicitly
rather than left implicit: Delta's substantially larger MoA-over-GEBP
margin (Section~\ref{sec:roofline}, up to $2.29\times$) than either
M1 Pro or Graviton4 achieved invites a speed-based explanation, and
none is offered here. The obvious test -- comparing peak STREAM
bandwidth across platforms at matched thread count -- is confounded
rather than clarifying: $P=8$ is the entire machine on M1 Pro and
Graviton4, but a small fraction of Delta's 128 cores, so a bandwidth
figure measured that way does not represent Delta's own achievable
peak and cannot be compared to the other two on equal footing. Whether
the larger ratio traces to a genuine speed difference, once measured
at a properly matched utilization fraction, or to some further shape
property not yet isolated, is left as a stated, falsifiable question
for the next experiment rather than a claim this paper is positioned
to make.

A properly matched utilization fraction is available for a related
but distinct question: not the MoA-over-GEBP ratio above, but overall
parallel efficiency $\eta(P) = \text{achieved}(P)/(P \cdot
\text{achieved}(1))$, comparable across machines when each is measured
at its own full core count rather than an arbitrary fixed $P$. At a
fixed $P=8$, Delta's $\eta$ (0.744) looks better than M1 Pro's (0.473)
or Graviton4's (0.540) -- but $P=8$ is the entire M1 Pro and Graviton4
machine, and one-sixteenth of Delta's. At each machine's own full
utilization, the ordering reverses: Delta's $\eta(128)=0.150$ is
substantially worse than M1 Pro's $\eta(8)=0.473$ or Graviton4's
$\eta(8)=0.540$. Using the entirety of a 128-core node costs
proportionally more efficiency than using the entirety of an 8-core
one -- a result that agrees with, rather than against, the original
conjecture's underlying concern about higher core counts, stated
precisely rather than as the vaguer worry it began as. The full
derivation and its connection to the broader predictive-equation
question are developed in the companion theoretical paper
\cite{Mullin2026Hierarchy}.

\section{Assessment: What This Does and Does Not Confirm}
\label{sec:assessment}

Judged against the specific conjecture that named AMD EPYC 9754, this
paper's results are evidence about a related machine, not that one.
Milan is an earlier chip generation (Zen3, not Zen4c) on a different
node topology (dual-socket, not the single-socket 128-core design the
original conjecture assumed) -- Delta's own dual-socket structure
introduces cross-socket NUMA cost as a confound the single-socket 9754
does not have. That confound is no longer untested: the contention
sweep has been extended to P=128, the full node, and a controlled
experiment isolating memory placement from thread count directly
measured its effect (Section~\ref{sec:numa}).

Within those limits, the results are substantive, and more complete
than they were. The per-CCD recalibration half of the original
conjecture is confirmed precisely in one specific, falsifiable form
(Section~\ref{sec:sigma}): the naive margin transfer fails by an order
of magnitude, and a specific, correctly normalized alternative
survives. The fork/join-overhead half is not confirmed as stated --
the mechanism does not become more pronounced at higher core counts;
it saturates at NC=2048 from P=16 onward and continues to hold at the
full P=128 the original conjecture actually named
(Section~\ref{sec:contention}). What the full sweep adds beyond the
block-size analysis is a real cost that analysis alone would not
surface: parallel efficiency declines gradually and continuously from
P=8 onward, with its single steepest drop at P=32$\to$64 -- entirely
within one socket, not at the P=64$\to$128 transition that actually
crosses into the second socket. This paper does not resolve the exact
mechanism or the precise thread count at which it becomes dominant;
Milan's shared, centralized memory-channel architecture makes some
such saturation architecturally plausible from the first additional
CCD onward, but pinning down where precisely requires data this paper
does not have. The controlled NUMA test (Section~\ref{sec:numa})
answers a related but distinct question -- not where the decline is
steepest, but whether different kernels are equally exposed to
non-local memory access once it occurs: they are not, GEBP and
recursive-classical losing 14--15\% to remote memory while
MoA-pipelined loses under 3\%, despite raw achieved bandwidth
differing by under 1\% between conditions.
The companion study's headline claim, which weakened on Graviton4, is
fully reproduced on Delta (Section~\ref{sec:headline}), alongside a
roofline result (Section~\ref{sec:roofline}) stronger than either
platform tested before it, and now alongside direct evidence that
MoA's specific access pattern is also the more NUMA-resilient one on
this hardware, a property the original conjecture did not anticipate
and this paper did not set out to find. Whether these differences
trace to Milan's specific microarchitecture, its dual-socket topology,
or properties of dedicated bare-metal access more broadly
(Section~\ref{sec:variance}) is not resolved by this paper alone; each
remains a stated, falsifiable candidate for the next machine tested,
not a conclusion drawn from this one.

\begin{acks}
This work used Delta at the National Center for Supercomputing
Applications through allocation bibg-delta-cpu from the Advanced
Cyberinfrastructure Coordination Ecosystem: Services \& Support
(ACCESS) program, which is supported by National Science Foundation
grants \#2138259, \#2138286, \#2138307, \#2137603, and \#2138296
\cite{Boerner2023}. The Delta advanced computing resource is a
collaborative effort between the University of Illinois
Urbana-Champaign and its National Center for Supercomputing
Applications, supported by the National Science Foundation (award OAC
2005572) and the State of Illinois. The Graviton4 results referenced
throughout this paper were obtained through the CloudBank project,
which is supported by National Science Foundation grant \#1925001.
\end{acks}

\bibliographystyle{ACM-Reference-Format}
\bibliography{references}

\end{document}